\documentclass[colorlinks,linkcolor=blue]{kybernetika}

\hidedoi
\setheader{}

\renewenvironment{Proof}[1][\Proofname]{%
\medskip\noindent#1\enspace\ignorespaces}{\hfill$\square$}

\usepackage[utf8]{inputenc}

\newtheorem{theorem}{Theorem}[section]
\newtheorem{lemma}[theorem]{Lemma}
\newtheorem{proposition}[theorem]{Proposition}
\newtheorem{corollary}[theorem]{Corollary}

\newtheorem{example}[theorem]{Example}

\newtheorem{problem}[theorem]{Problem}
\newtheorem{conjecture}[theorem]{Conjecture}

\newcommand{\Rb}{\mathbb{R}}
\newcommand{\Eb}{\mathbb{E}}

\newcommand{\Ecal}{\mathcal{E}}
\newcommand{\Fcal}{\mathcal{F}}
\newcommand{\Ncal}{\mathcal{N}}
\newcommand{\Pcal}{\mathcal{P}}

\newcommand{\ol}{\overline}

\newcommand{\Bbar}{\overline B}

\newcommand{\<}{\langle}
\renewcommand{\>}{\rangle}

\DeclareMathOperator{\cs}{cs}  
\DeclareMathOperator{\conv}{conv}

\DeclareMathOperator{\relint}{ri}
\DeclareMathOperator{\supp}{s}

\renewcommand{\theta}{\vartheta}

\begin{document}
\pagestyle{myheadings}

\title{Maximizing the Bregman divergence\texorpdfstring{\\}{ }from a Bregman family}

\author{Johannes Rauh and Franti\v sek Mat\'u\v s}

\contact{Johannes}{Rauh}{Max Planck Institute for Mathematics in the Sciences, Inselstra\ss e 22, 04103 Leipzig, Germany}{jrauh@mis.mpg.de}
\markboth{J.~Rauh and F.~Mat\'u\v s}{Maximizing the Bregman divergence}

\maketitle

\begin{abstract}
  The problem to maximize the information divergence from an exponential family is generalized to the setting of Bregman divergences and suitably defined Bregman families.
\end{abstract}


\tableofcontents{}

\section{Introduction}

Let $Z$ be a finite set.  Denote by $\Pcal(Z)$ the set of probability measures (pm's) with support contained in~$Z$.
Let $\Ecal\subseteq\Pcal(Z)$ be an exponential family supported on~$Z$.
For $P, Q\in\Pcal(Z)$ denote by $D(P\|Q)$ the \emph{information divergence} (also known as \emph{Kullback-Leibler divergence}), and let $D(P\|\Ecal) \triangleq \inf_{Q\in\Ecal}D(P\|Q)$.
In 2002, Nihat Ay formulated the following optimization problem~\cite{Ay02:Pragmatic_structuring}:
\begin{problem}
  \label{prob:main-problem-KL}
  Maximize $D(P\|\Ecal)$ over all probability distributions $P$ on~$Z$.
\end{problem}
The original motivation came from theoretical studies of the infomax principle.
Insight into this problem can also be used to bound approximation errors of machine learning models or other statistical models~\cite{MontufarRauhAy11:Expressive_Power_and_Approximation_Errors_of_RBMs,MontufarRauhAy13:Maximal_KL_from_network_models}.

Since 2002, progress has been made in different directions.  The problem was attacked for particular classes of exponential families, with a particular focus on hierarchical models~\cite{MatusAy03:On_Maximization_of_the_Information_Divergence,Matus04:Maximization_from_binary_iid_seqs,AyKnauf06:Maximizing_Multiinformation,Matus09:Divergence_from_factorizable_distributions}.  A full characterization of the first order optimality conditions was given in~\cite{Matus07:Optimality_conditions}.

In 2010, the first author found a surprising connection to another optimization problem~\cite{Rauh11:Finding_Maximizers}: Let $A$ be
the \emph{design matrix} (or \emph{sufficient statistics matrix}) of~$\Ecal$, where the columns of $A$ are indexed by~$Z$.
Any $u\in\ker A$ can
be written uniquely as a difference $u = u^{+}-u^{-}$ of non-negative vectors $u^{ +},u^{-}$ of
disjoint support.  For $u\in\ker A\setminus\{0\}$ with $\sum_{x\in Z}u^{+}(x) = \sum_{x\in Z}u^{-}(x) = 1$ let
\begin{equation*}
  \ol D(u) = H(u^{-}) - H(u^{+}) = \sum_{x\in Z}u(x)\log|u(x)|,
\end{equation*}
where $H$ denotes the (Shannon) entropy.
The second optimization problem is:
\begin{problem}
  \label{prob:bar-problem-KL}
  Maximize $\ol D(u)$ over the set all $u\in\ker A$ that satisfy $\sum_{x\in Z}u^{+}(x) = \sum_{x\in Z}u^{-}(x) = 1$.
\end{problem}
The optimization problem~\ref{prob:bar-problem-KL} is easier than the optimization
problem~\ref{prob:main-problem-KL}, since the function to be optimized in~\ref{prob:main-problem-KL}
is itself defined by an optimization problem.

Both authors showed in~\cite{MatusRauh11:Maximization-ISIT2011} that the map $u\mapsto u^{+}$ induces a one-to-one correspondence between the points that satisfy the respective critical equations of~\ref{prob:bar-problem-KL} and~\ref{prob:main-problem-KL}, and that this correspondence restricts to bijections of the sets of local optimizers and global optimizers, respectively.

The authors found this connection quite surprising.  To better understand this result, the second author suggested to try to generalize the result to the setting of Bregman divergences and Bregman families.
The present paper summarizes the results of this investigation.

The first step is the definition of a function $\Bbar$ that serves as an analogue
of $\ol D$ in the general case.  Once this definition is in place, the equivalence of the global
maximizers is rather straightforward (Theorem~\ref{thm:equivalence}).  What makes the general
Bregman case more difficult is that $\Bbar$ is only defined implicity as a solution of an
optimization problem.  Hence, the criticality conditions of $\Bbar$ are currently unknown.
If the optimization problem underlying $\Bbar$ always has a unique solution
(Conjecture~\ref{con:uniqueness-codim-one}), then the bijection of the local maximizers also
generalizes (Theorem~\ref{thm:local-maxi}).

Section~\ref{sec:Bregman-setting} recalls definitions and basic properties of Bregman divergences and introduces Bregman families.  Section~\ref{sec:maxim-bregm-diverg} discusses the problem of maximizing the Bregman divergence from a Bregman family.  Section~\ref{sec:Bbar} introduces the function $\Bbar$ that corresponds to the function~$\ol D$.  Section~\ref{sec:equivalence} contains the main results that relates the problems to maximize the Bregman divergence and $\Bbar$, respectively.  Section~\ref{sec:classical} compares the results to the results of~\cite{MatusRauh11:Maximization-ISIT2011} that concern the classical case of exponential families and the information divergence.

\section{Preliminaries: Bregman divergences and Bregman families}
\label{sec:Bregman-setting}

This section summarizes the relevant results about Bregman divergences and Bregman families.  The end of the section contains in Example~\ref{ex:classical-case} the special case of information divergence and exponential families.
For more details and generalizations to the case where $Z$ is not finite see~\cite{MatusCsiszar12:Bregman_Pythagorean_identities}.

It is wellknown that one can associate to each exponential family a Bregman divergence by expressing the information divergence within the exponential family in terms of the exponential family's natural parameters.  However, this construction is not used in this paper.  Instead, starting from a particular Bregman divergence, a family of distributions is defined, called a \emph{Bregman family}.  These Bregman families generalize exponential families.

Consider a finite set $Z$.  For each $z\in Z$ let $\beta_{z}:(0,+\infty)\to\Rb$ be a convex
differentiable function with $\lim_{x\to 0+}\beta_z'(x) = -\infty$ and
$\lim_{x\to+\infty}\beta_z'(x) = +\infty$, where $\beta'_{z}(x)$ denotes the derivative of
$\beta_{z}(x)$ with respect to~$x$.  Then the convex conjugate
(see~\cite{Rockafellar70:Convex_Analysis}) 
\begin{equation*}
  \beta_z^{*}(t) = \sup_{x}\big\{tx - \beta_{z}(x)\big\}
\end{equation*}
is differentiable and ranges
between $-\lim_{x\to 0+}\beta_z(x)$ and $+\infty$.  The derivative
$e_{z}(x) \triangleq \beta_z^{*\prime}(x)$ is continuous and strictly increases from 0 to $+\infty$.
Therefore, the inverse function $l_{z}(y) \triangleq e_{z}^{-1}(y)$ exists for $0<y<+\infty$, is
continuous and strictly increases from $-\infty$ to~$+\infty$.  The inverse function satisfies
$l_{z}(y)=\beta'_{z}(y)$.

The following lemma is a standard result in convex analysis (see~\cite{Rockafellar70:Convex_Analysis} or Lemma~2.2
in~\cite{MatusCsiszar12:Bregman_Pythagorean_identities}):
\begin{lemma}
  \label{lem:2.2}
  $\beta_{z}(e_{z}(r)) = r e_{z}(r) - \beta^{*}(r)$ for all $r < \beta'(+\infty)$.
\end{lemma}

Consider a function~$f:Z\to\Rb^{d}$.  For $\theta\in\Rb^{d}$ define a pm
$P_{\theta}:z\mapsto e_{z}(\<\theta,f(z)\> - \Lambda(\theta))$, where $\Lambda(\theta)$ is the
unique solution of $\sum_{z\in Z}e_{z}(\<\theta,f(z)\> - r)=1$ in~$r$.  The subset
\begin{equation*}
  \Ecal = \Ecal_{f} \triangleq \{ P_{\theta} : \theta\in\Rb^{d} \}
\end{equation*}
of $\Pcal(Z)$ will be called a \emph{Bregman family} in the following.\footnote{The second author
  had originally given the name \emph{generalized exponential family} to~$\Ecal$, which is also used
  by other authors.  However, since that name is not very specific and since there are many
  different ways in which exponential families can be generalized, this paper now uses the name
  \emph{Bregman family}.}
The matrix $A$ with columns $f(z)$ for $z\in Z$ (after fixing an ordering of~$Z$) is called the \emph{design matrix} of~$\Ecal$.

The set $\cs(\Ecal)\triangleq \conv\{f(z) : z\in Z\}$ is called the
\emph{convex support} of~$\Ecal$.  The convex support is a (convex) polytope.  A set $S\subseteq Z$
is called \emph{facial} for $\Ecal$ if and only if $\conv\{f(z) : z\in S\}$ is a face of
$\cs(\Ecal)$.

The \emph{Bregman divergence} of $u,v: Z\to [0,+\infty)$ is
\begin{equation*}
  B(u,v) = \sum_{z\in Z}\big[
    \beta_{z}(u(z)) - \beta_{z}(v(z)) - \beta'_{z}(v(z))[u(z)-v(z)]
  \big].
\end{equation*}
The Bregman divergence of $P\in\Pcal(Z)$ from a Bregman family $\Ecal$ is
\begin{equation*}
  B(P,\Ecal) \triangleq \inf_{Q\in\Ecal} B(P,Q).
\end{equation*}

When the minimizer in the definition of $B(P,\Ecal)$ does not exist, one can find a minimizer in the closure $\ol\Ecal$ of~$\Ecal$, where the closure can be taken with respect to the canonical topology on the finite dimensional convex polytope $\Pcal(Z)$.
Just as in the classical case of an exponential family, one can prove the following statements:
\begin{proposition}
  \label{prop:projection-E}
  Let $\Ecal$ be a Bregman family.
  \begin{enumerate}
  \item For any $P\in\Pcal(Z)$ there exists a unique pm $\Pi_{\Ecal,P}\in\ol\Ecal$ with
    \begin{equation*}
      B(P,\Pi_{\Ecal,P}) = B(P,\Ecal).
    \end{equation*}
  \item Let $P\in\Pcal(Z)$ and $Q\in\ol\Ecal$.  If $\Eb_{P}[f] = \Eb_{Q}[f]$, then $Q = \Pi_{\Ecal,P}$.
  \item 
    Let $P\in\Pcal(Z)$.  The unique global minimum of $H(Q)\triangleq \sum_{z\in Z}\beta_{z}(Q(z))$
    for pm's $Q\in\Pcal(Z)$ with $\Eb_{P}[f] = \Eb_{Q}[f]$ is given by $Q = \Pi_{\Ecal,P}$.
  \item The support $\supp(\Pi_{\Ecal,P})$ is the smallest facial set containing $\supp(P)$.
  \end{enumerate}
  The pm $\Pi_{\Ecal,P}$ is called the \emph{generalized reverse Bregman projection} ($rB$-projection) of $P$ to~$\Ecal$.
  Here, ``generalized'' may be dropped whenever $\Pi_{\Ecal,P}\in\Ecal$.
  If the Bregman family $\Ecal$ is clear from the context, $\Pi_{\Ecal,P}$ is abbreviated by~$\Pi_{P}$.
\end{proposition}
\begin{proposition}
  \label{prop:closure-E}
  Let $\Ecal$ be a Bregman family.
  \begin{enumerate}
  \item The map $\mu:P\in\Pcal(Z)\mapsto\Eb_{P}[f]$ surjects onto $\cs(\Ecal)$.  It restricts
    to a homeomorphism $\ol\Ecal\cong\cs(\Ecal)$.
  \item $\ol\Ecal = \bigcup_{F} \Ecal_{F}$, where $F$ runs over all sets $F\subseteq Z$ that are
    facial with respect to~$\Ecal$ and where $\Ecal_{F}$ is the Bregman family defined on~$F$ using
    $f|_{F}$.
  \end{enumerate}
\end{proposition}

For exponential families, the statements in Propositions~\ref{prop:projection-E}
and~\ref{prop:closure-E} are well-known and go back at least
to~\cite{Barndorff78:Information_and_Exponential_Families}.  The statements continue to hold for
exponential families when $Z$ is replaced by a more general measure spaces~$Z$, as studied
in~\cite{CsiszarMatus05:Closures_of_exp_fam,CsiszarMatus08:GMLE_for_Exp_Fam}.  The extended arXiv
version of~\cite{WangRauhMassam19:Approximating_faces_w_arxiv} contains a
direct proof of the discrete case, which relies on algebraic insights
from~\cite{GeigerMeekSturmfels06:Toric_Algebra_Graphical_Models}.

\smallskip

For a distribution of the form~$Q:z\mapsto e_{z}(r_{z})$,
with $r_{z}\in\Rb$, by Lemma~\ref{lem:2.2},
\begin{multline*}
  B(P,Q) = \sum_{z\in Z}\big[
    \beta_{z}(P(z)) - \beta_{z}(e_{z}(r_{z})) - r_{z}[P(z)-Q(z)]
  \big] \\
  = \sum_{z\in Z}\big[
  \beta_{z}(P(z)) - r_{z} e_{z}(r_{z}) + \beta^{*}(r_{z}) - r_{z}[P(z)-Q(z)]
  \big] \\
  = \sum_{z\in Z}\big[
  \beta_{z}(P(z)) + \beta^{*}(r_{z}) - r_{z} P(z)
  \big].
\end{multline*}
When $Q\in\Ecal$, then $r_{z}$ is of the the form $\<\theta,f(z)\> - \Lambda(\theta)$.  Thus,
\begin{multline}
  \label{eq:B-Phi}
  B(P,\Ecal)
  = \sum_{z}\beta_{z}(P(z)) - \sup_{\theta}\Big[
    \<\theta, \sum_{z}f(z) P(z)\> - \Lambda(\theta) - \sum_{z}\beta_{z}^{*}(\<\theta,f(z)\> - \Lambda(\theta))
  \Big] \\
  = \sum_{z}\beta_{z}(P(z)) - \sup_{\theta}\Big[ \<\theta, \mu(P)\> - \Upsilon(\theta) \Big],
\end{multline}
where $\Upsilon(\theta) = \Lambda(\theta) + \sum_{z}\beta_{z}^{*}(\<\theta,f(z)\> - \Lambda(\theta))$.
\begin{theorem} 
  $\Upsilon$ is convex.
  Its partial derivatives are
  \begin{equation*}
    \frac{\partial}{\partial\theta_{i}} \Upsilon(\theta)
    = \Eb_{\theta}[f_{i}] = \mu(P_{\theta})_{i},
  \end{equation*}
  where $\Eb_{\theta}$ denotes the expected value taken with respect to~$P_{\theta}$.
  The map $\nabla\Upsilon:\Rb^{d}\to\cs(\Ecal)$ is surjective.
  The Hessian of $\Upsilon$ is positive definite.
\end{theorem}
\begin{Proof}
  \begin{align*}
    \frac{\partial}{\partial\theta_{i}} \Upsilon(\theta)
    &= \partial_{i} \Lambda(\theta)
      + \sum_{z} \beta^{*\prime}_{z}(\<\theta,f(z)\> - \Lambda(\theta)) [f_{i}(z) - \partial_{i} \Lambda(\theta)] \\
    &= \sum_{z} \beta^{*\prime}_{z}(\<\theta,f(z)\> - \Lambda(\theta)) f_{i}(z), \\
    \frac{\partial^{2}}{\partial\theta_{i}\partial\theta_{j}} \Upsilon(\theta)
    &= \sum_{z} \beta^{*\prime\prime}_{z}(\<\theta,f(z)\> - \Lambda(\theta)) f_{i}(z) [f_{j}(z) - \partial_{j}\Lambda(\theta)] \\
    & = \sum_{z} \beta^{*\prime\prime}_{z}(\<\theta,f(z)\> - \Lambda(\theta)) [f_{i}(z) - \partial_{i}\Lambda(\theta)][f_{j}(z) - \partial_{j}\Lambda(\theta)] \succeq 0,
  \end{align*}
  where the last equality follows from deriving the defining equation
  $\sum_{z}\beta^{*\prime}_{z}(\<\theta,f(z)\> - \Lambda(\theta))=1$ of~$\Lambda(\theta)$:
  \begin{equation*}
    0 = \frac{\partial}{\partial\theta_{j}} \sum_{z}\beta^{*\prime}_{z}(\<\theta,f(z)\> - \Lambda(\theta))
    = \sum_{z}\beta^{*\prime\prime}_{z}(\<\theta,f(z)\> - \Lambda(\theta)) [f_{j}(z) - \frac{\partial}{\partial\theta_{j}} \Lambda(\theta)].
  \end{equation*}
  This shows convexity.

  It is clear that $\Eb_{\theta}[f]=\mu(P_{\theta})$ belongs to $\conv\{f(z) : z\in Z\}$.
  Surjectivity follows from Proposition~\ref{prop:closure-E}.
\end{Proof}

It follows from the properties of convex conjugation:
\begin{corollary}
  \label{cor:Nabla-Upsilon}
  The maps $\theta\mapsto\nabla\Upsilon(\theta)$ and $\mu\mapsto\nabla\Upsilon^{*}(\mu)$ are mutual
  inverses in the relative interiors of their respective domains.
  If $\Pi_{P}\in\relint(\Ecal)$, then
  $\Pi_{P}=P_{\theta}$ for $\theta=\nabla\Upsilon^{*}(\mu(P))$.
\end{corollary}

Let $H(P) = \sum_{z\in Z}\beta_{z}(P(z))$ as in Proposition\ref{prop:projection-E}.  Then~\eqref{eq:B-Phi} rewrites to
\begin{equation*}
  B(P,\Ecal) = H(P) - \Upsilon^{*}(\mu(P)),
\end{equation*}
where $\Upsilon^{*}$ denotes the convex conjugate of~$\Upsilon$.
From this equality follows the next result, which can also be seen as a kind of Pythagorean identity:
\begin{corollary}
  \label{cor:triangleH}
  $B(P,\Ecal) = H(P) - H(\Pi_{P})$ for all~$P\in\Pcal(Z)$.
\end{corollary}
\begin{Proof}
  $B(P,\Ecal) = B(P,\Ecal) - B(\Pi_{P},\Ecal)
  = H(P) - H(\Pi_{P})$, since $\mu(P) = \mu(\Pi_{P})$. 
\end{Proof}

\begin{example}
  \label{ex:classical-case}
  Let 
  $\beta_{z}(x)= x\ln(x/\nu(z)) + x$ for all~$z\in Z$.  Then $\beta^{*}_{z}(x) = \nu(z)\exp(x)$ and
  $l_{z}(x) = \beta'_{z}(x) = \ln(x/\nu(z))$, and so $e_{z}(x) = \nu(z)\exp(x)$.  In this case,
  $\Ecal$ is an exponential family with reference measure~$\nu$, and $B$ equals the information
  divergence.  Since $\beta^{*}_{z}=e_{z}(x)$, it follows that
  $\sum_{z}\beta_{z}^{*}(\<\theta,f(z)\> - \Lambda(\theta)) = 1$.  Therefore,
  $\Upsilon(\theta) = 1 + \Lambda(\theta)$.  In the classical case, $\Lambda$ is called the \emph{partition function}, and convexity of $\Lambda$ is well-known
  and widely used.  In the general case, $\Lambda$ itself need not be convex.
\end{example}

\section{Maximizing the Bregman divergence from a Bregman family}
\label{sec:maxim-bregm-diverg}

Let $\Ecal$ be a Bregman family.
The following problem generalizes Problem~\ref{prob:main-problem-KL}:
\begin{problem}
  \label{prob:B}
  Maximize $B(P,\Ecal)$ over $P\in\Pcal(Z)$.
\end{problem}
\begin{theorem}
  \label{thm:projection-property}
  If $P\in\Pcal(Z)$ is a local maximizer of $B(\cdot,\Ecal)$, then the map $z\mapsto l(P(z)) - l(\Pi_{P}(z))$ is constant for $z\in\supp(P)$
\end{theorem}
\begin{Proof}
  If $\mu(P) = \sum_{z}f(z)P(z)$ does not lie in the relative interior of $\cs(\Ecal)$, by
  Proposition~\ref{prop:closure-E}, one may replace $\Ecal$ by $\Ecal_{F}$ for some
  suitable~$F\subsetneq Z$.  Thus, without loss of generality, assume that
  $\mu(P)$ lies in the relative interior of $\cs(\Ecal)$.

  Let $w\in\Rb^{Z}$ with $\sum_{z}w(z) = 0$ and $\supp(w)\subseteq\supp(P)$.  For $\epsilon>0$ small,
  \begin{multline*}
    B(P+\epsilon w,\Ecal) \approx
    H(P) + \epsilon \sum_{z}\beta_{z}'(P(z)) w(z) \\
    - \Upsilon^{*}\big(\mu(P)\big) - \epsilon \Big\< \nabla\Upsilon^{*}\big(\mu(P)\big), \sum_{z}f(z) w(z) \Big\>
  \end{multline*}
  to first order in~$\epsilon$.
  Let $\theta=\nabla\Upsilon^{*}\big(\mu(P)\big)$.
  Then $\Pi_P = P_\theta$ by Corollary~\ref{cor:Nabla-Upsilon}, and
  \begin{equation*}
    \< \theta, \sum_{z}f(z)w(z) \>
    = \sum_{z} [\<\theta, f(z)\> - \Lambda(\theta) ] w(z)
    = \sum_{z} \beta'_{z}(\Pi_{P}(z))w(z),
  \end{equation*}
  since $\beta'$ and $\beta^{*\prime}$ are mutual inverses to each other.  In total,
  \begin{equation*}
    B(P+\epsilon w,\Ecal) \approx
    H(P) - \Upsilon^{*}\big(\mu(P)\big) + \epsilon \sum_{z}\big[\beta_{z}'(P(z)) - \beta'_{z}(\Pi_{P}(z))\big] w(z),
  \end{equation*}
  whence $\sum_{z}\big[\beta_{z}'(P(z)) - \beta'_{z}(\Pi_{P}(z))\big] w(z) = 0$ if $P$ is a critical
  point.  This equality holds for all~$w\in\Rb^{Z}$ with $\sum_{z}w(z) = 0$ and
  $\supp(w)\subseteq\supp(P)$.  Therefore, $\beta_{z}'(P(z)) - \beta'_{z}(\Pi_{P}(z))$ is constant
  for $z\in\supp(P)$.
\end{Proof}

\begin{corollary}
  Let $P\in\Pcal(Z)$ be a local maximizer of $B(\cdot,\Ecal)$, and let $u=P-\Pi_{P}$.  Then $\supp(u^{+})=\supp(P)$.

  If $\beta_{x}=\beta_{y}$ for $x,y\in\supp(P)$, then $u^{+}(x)\ge u^{+}(y)$ if and only if
  $P(x)\ge P(y)$.
\end{corollary}
\begin{Proof}
  By Theorem~\ref{thm:projection-property}, there exists a constant $c$ such that $l(P(z)) - l(\Pi_{P}(z))=c$ for $z\in s(P)$.  The number $c$ equals
  the unique solution of the equation
  \begin{equation*}
    \sum_{z\in\supp(P)}e_{z}(l_{z}(\Pi_{P}(z)) + c) = \sum_{z\in\supp(P)}P(z) = 1.
  \end{equation*}
  Since all functions $e_{z}$ are increasing, $c>0$.  Thus, if $z\in\supp(P)$, then
  $l_{z}(P(z)) > l_{z}(\Pi_{P}(z))$, and so $P(z)>\Pi_{P}(z)$.  This implies
  $\supp(P)\subseteq\supp(u^{+})$.  On the other hand, if $z\notin\supp(P)$, then
  $\Pi_{P}(z)\ge P(z)$, and so $u(z)\le 0$, which implies $\supp(P)\supseteq\supp(u^{+})$.
\end{Proof}

As in the classical case, one shows~\cite{MatusAy03:On_Maximization_of_the_Information_Divergence}:
\begin{proposition}
  Any $P\in\Pcal(Z)$ that globally maximizes $B(\cdot,\Ecal)$ satisfies $|\supp(P)|\le\dim(\Ecal) + 1$.
\end{proposition}

\section{The function \texorpdfstring{$\Bbar$}{Bbar} and the alternative optimization problem}
\label{sec:Bbar}

For each real vector-valued function $f:Z\to\Rb^{d}$ let
\begin{equation*}
  \Ncal = \Ncal(f) = \Big\{u\in\Rb^{Z} : \sum_{z\in Z}f(z)u(z)=0\text{ and }\sum_{z\in Z}u(z)=0\Big\}.
\end{equation*}
If $A$ is a design matrix, then $\Ncal=\big\{u\in\ker A:\sum_{z\in Z}u(z)=0\big\}$.

Let $u:Z\to\Rb$ be a real function satisfying $\sum_{x\in Z}u(x)=0$.  To each such $u$ associate a
function $f_{u}$ such that $\Ncal(f_{u}) = \Rb u$, and let $\Fcal_{u}=\Ecal_{f_{u}}$.  Then
$\Fcal_{u}$ has codimension one.
By Proposition~\ref{prop:projection-E}, the difference $P-\Pi_{\Fcal_{u},P}$ lies in~$\Rb u$.
\begin{lemma}
  \label{lem:P-u}
  Let $P\in\Pcal(Z)$, and let $u=P-\Pi_{P}$.  Then $\Pi_{\Fcal_{u},P}=\Pi_{P}$.
\end{lemma}
\begin{Proof}
  From $\Ecal\subseteq\Fcal_{u}$ follows $\Pi_{P}\in\Fcal_{u}$.  Together with $P-\Pi_{P}=u$, the
  statement follows from Proposition~\ref{prop:projection-E}.
\end{Proof}

\bigskip
$\Pcal(Z)$ can be partitioned into $\Pcal_{u}^{+}\cup\ol\Fcal_{u}\cup\Pcal_{u}^{-}$, where
\begin{equation*}
  \Pcal_{u}^{+}=\big\{ P\in\Pcal(Z) : \<P-\Pi_{\Fcal_{u},P},u\> > 0 \big\}, \quad \Pcal_{u}^{-}=\big\{ P\in\Pcal(Z) : \<P-\Pi_{\Fcal_{u},P},u\> < 0 \big\}.
\end{equation*}
The definition and Lemma~\ref{lem:P-u} imply:
\begin{lemma}
  $P \in \Pcal_{P-\Pi_{P}}^{+}$ for any $P\in\Pcal(Z)\setminus\ol\Ecal$.
\end{lemma}

In the classical case, the maximizer of the information divergence from an arbitrary exponential family $\Ecal$ need not be unique~\cite{MatusAy03:On_Maximization_of_the_Information_Divergence}.  However, when $\Ecal=\Fcal_{u}$ has codimension one, there are precisely two local maximizers $u^{+}$ and $u^{-}$, one on each side of~$\Ecal$~\cite[Section~VI]{Rauh11:Thesis}.  This motivates the following conjecture:
\begin{conjecture}
  \label{con:uniqueness-codim-one}
  The map $P\in\Pcal_{u}^{+}\mapsto B(P,\Fcal_{u})$ has a unique local (and global) maximizer.
\end{conjecture}
The proof of the conjecture in the classical case relies on applying properties of the logarithm to
the criticality conditions in Theorem~\ref{thm:projection-property}.  It is not possible to apply
this proof to the general case of the conjecture.

For any function $u:Z\to\Rb$ that satisfies $\sum_{z\in Z}u(z) = 0$ let
\begin{equation*}
  \Bbar(u) \triangleq \max \big\{ B(P,\Fcal_{u}) : P\in\ol\Pcal_{u}^{+} \big\},
\end{equation*}
where $\ol\Pcal_{u}^{+}=\Pcal_{u}\cup\ol\Fcal_{u}$ denotes the closure of~$\Pcal_{u}^{+}$.  The map
$\Bbar$ is continuous and welldefined since $\ol\Pcal_{u}^{+}=\Pcal_{u}^{+}\bigcup\ol\Fcal_{u}$ is compact.
If $u\neq 0$, then this maximum lies in $\Pcal_{u}^{+}$, and $\ol B(u) > 0$.
The function $\Bbar$ satisfies $\Bbar(\lambda u) = \Bbar(u)$ for all~$\lambda>0$.

\begin{problem}
  \label{prob:Bbar}
  Maximize the function $u\in\Ncal\setminus\{0\}\mapsto\Bbar(u)$.
\end{problem}

The intuition behind the definition of $\Bbar$ and Problem~\ref{prob:Bbar} is the following: instead
of directly searching for a maximizer $P$ of~$B(\cdot,\Ecal)$, one may try to determine the vector
$u=P-\Pi_{P}$, which can be seen as a direction within the probability simplex.  Thus, the task is
to find a direction in which it is possible to achieve large values of~$B(\cdot,\Ecal)$.
When analyzing the direction~$u$, Lemma~\ref{lem:P-u} says that one may just as well reaplace $\Ecal$
by~$\Fcal_{u}$.


\section{Equivalence of the maximizers}
\label{sec:equivalence}

The following theorem specifies the relations between the problems~\ref{prob:B} and~\ref{prob:Bbar}.
It corresponds to \cite[Theorem~3]{Rauh11:Finding_Maximizers}.
\begin{theorem}
  \label{thm:equivalence}
  \begin{enumerate}
  \item $\max_{P\in\Pcal(Z)} B(P,\Ecal) = \max_{u\in\Ncal\setminus\{0\}}\Bbar(u)$.
  \item If $P$ is a global maximizer of problem~\ref{prob:B}, then $P-\Pi_{P}$ is a global maximizer
    of problem~\ref{prob:Bbar}, and $B(P,\Ecal)=\Bbar(P-\Pi_{P})$.
  \item If $u$ is a global maximizer of problem~\ref{prob:Bbar} and if $\Bbar(u) = B(P,\Fcal_{u})$,
    then $P$ is a global maximizer of problem~\ref{prob:B}, and $\Bbar(u) = B(P,\Ecal)$.
\end{enumerate}
\end{theorem}

The proof of Theorem~\ref{thm:equivalence} is based on the following auxilliary theorem, which corresponds to \cite[Theorem~2]{MatusRauh11:Maximization-ISIT2011}.
\begin{theorem}
  \label{thm:Bbar_B_ineq}
  $\Bbar(P - \Pi_{P}) \ge B(P,\Ecal)$ for any $P\in\Pcal(Z)\setminus\ol\Ecal$.  If
  $u\in\Ncal\setminus\{0\}$ and $P\in\Pcal(Z)$ satisfy $\Bbar(u) = B(P,\Fcal_{u})$, then
  $B(P,\Ecal) \ge \Bbar(u)$, with equality if and only if $P - \Pi_{P}=\lambda u$ for
  some~$\lambda>0$.
\end{theorem}
\begin{Proof}
  The first statement follows from Lemma~\ref{lem:P-u}, as $\Bbar(P - \Pi_{P}) \ge B(P,\Fcal_{P-\Pi_{P}}) = B(P,\Ecal)$.
  For the second statement observe that
  from $\Ecal\subseteq\Fcal_{u}$ follows $B(P,\Ecal)\ge B(P,\Fcal_{u}) = \Bbar(u)$.
\end{Proof}

Theorem~\ref{thm:equivalence} follows directly from Theorem~\ref{thm:Bbar_B_ineq}.

In~\cite[Theorem~1]{MatusRauh11:Maximization-ISIT2011} it was shown that the points that satisfy the
respective critical equations (i.e.\ the equality conditions among the first order conditions) of the two problems~\ref{prob:B} and~\ref{prob:Bbar} and the local maximizers of the two problems are
also in one-to-one correspondence in the classical case.  Discussing the criticality conditions is difficult, as no explicit formula for $\Bbar$ is known, and if Conjecture~\ref{con:uniqueness-codim-one} is wrong, it is improbable that $\Bbar$ is differentiable.  If the conjecture is true, one can at least prove that the local maximizers of the two problems are related, as Theorem~\ref{thm:local-maxi} below will show.

Assume that Conjecture~\ref{con:uniqueness-codim-one} is true, and let
$\Phi(u)\triangleq\arg\max_{Q\in\Pcal_{u}^{+}(Z)}B(Q,\Fcal_{u})$ for $u\in\Ncal\setminus\{0\}$.  By
assumption, $\Phi$ is well-defined and continuous.  The map
$\Psi:\Pcal\to\Ncal,P\mapsto P-\Pi_{P}$ is also continuous.  With
these two maps, Theorem~\ref{thm:Bbar_B_ineq} can be reformulated as follows:
\begin{corollary}
  \label{cor:Bbar_B_ineq}
  If Conjecture~\ref{con:uniqueness-codim-one} is true, then:
  \begin{enumerate}
  \item $\Bbar(\Psi(P)) \ge B(P,\Ecal)$ for all~$P\in\Pcal(Z)$, with equality if and only if $P = \Phi(\Psi(P))$.
  \item $B(\Phi(u)) \ge \Bbar(u)$ for all~$u\in\Ncal\setminus\{0\}$, with equality if and only if
    $\Psi(\Phi(u)) = \lambda u$ for some $\lambda > 0$.
  \end{enumerate}
\end{corollary}
\begin{lemma}
  \label{lem:loc_max_E_F}
  \begin{enumerate}
  \item If $u\in\Ncal\setminus\{0\}$ is a local maximizer of~$\Bbar$, then
    $\Pi_{\Phi(u)}= \Pi_{\Fcal_{u},\Phi(u)}$.  Thus, if Conjecture~\ref{con:uniqueness-codim-one} is
    true, then $\Psi(\Phi(u)) = \lambda u$ for some $\lambda > 0$.
  \item If $P\in\Pcal(Z)$ is a local maximizer of~$B(\cdot,\Ecal)$, then $P$ is a local maximizer of
    $B(\cdot,\Fcal_{\Psi(P)})$.    Thus, if Conjecture~\ref{con:uniqueness-codim-one} is
    true, then $\Phi(\Psi(P)) = P$.
  \end{enumerate}
\end{lemma}
\begin{Proof}
  For the first statement, let $P=\Phi(u)$.  Suppose that $\Pi_{P}\neq\Pi_{\Fcal_{u},P}$,
  and let $Q$ be a pm in the convex hull of $\Pi_{P}$ and $\Pi_{\Fcal_{u},P}$.  Since $H$ is
  strictly convex and by Proposition~\ref{prop:projection-E}, $H(\Pi_{P}) < H(Q) < H(\Pi_{\Fcal_{u},P})$.
  Corollary~\ref{cor:Bbar_B_ineq} implies
  $\Bbar(P - Q) \ge B(P,\Fcal_{P-Q}) \ge H(P) - H(Q) > H(P) - H(\Pi_{\Fcal_{u},P}) = \Bbar(u)$.
  This contradicts the assumption that $u$ is a local maximizer.  Hence,
  $\Pi_{\Fcal_{u},P}=\Pi_{P}$, and $u = \Phi(P)$.

  For any $Q\in\Pcal(Z)$, if $B(Q,\Ecal)\le B(P,\Ecal)$, then
  $B(Q,\Fcal_{P-\Pi_{P}})\le B(Q,\Ecal) \le B(P,\Ecal) = B(P,\Fcal_{P-\Pi_{P}})$, where the last
  equality uses Lemma~\ref{lem:P-u}.  This proves the second statement.
\end{Proof}

\begin{theorem}
  \label{thm:local-maxi}
  Assume that Conjecture~\ref{con:uniqueness-codim-one} is true.
  If $P$ is a local maximizer of $B(P,\Ecal)$, then $\Psi(P)$ is a local maximizer of~$\Bbar$.
  If $u$ is a local maximizer of $\Bbar$, then $\Phi(u)$ is a local maximizer of~$B(P,\Ecal)$.
\end{theorem}
\begin{Proof}
  Let $P$ be a local maximizer of $B(P,\Ecal)$.  Let $U$ be a neighbourhood of $P$ in
  $\Pcal(Z)$ such that $B(Q,\Ecal)\le B(P,\Ecal)$ for all $Q\in U$.  Then $U':=\Phi^{-1}(U)$ is a
  neighbourhood of~$P$ by Lemma~\ref{lem:loc_max_E_F}.  If $v\in U'$, then
  Corollary~\ref{cor:Bbar_B_ineq} implies
  \begin{equation*}
    \Bbar(\Psi(P)) \ge B(P,\Ecal) \ge B(\Phi(v),\Ecal)
    = \Bbar(v).
  \end{equation*}
  for all~$v\in U'$.  This proves the first statement.

  Let $u\in\Ncal\setminus\{0\}$ be a local maximizer of~$\Bbar$.  Let $U'$ be a neighbourhood
  of $u$ in~$\Ncal\setminus\{0\}$ with $\ol B(u)\ge\ol B(v)$ for all~$v\in U'$.
  Then $U:=\Psi^{-1}(U')$ is a neighbourhood of~$P$ by Lemma~\ref{lem:loc_max_E_F}.  If
  $Q\in U$, then Corollary~\ref{cor:Bbar_B_ineq} implies
  \begin{equation*}
    B(\Phi(u),\Ecal) \ge \Bbar(u) \ge \Bbar(\Psi(Q)) \ge B(Q,\Ecal).
  \end{equation*}
  This proves the second statement.
\end{Proof}

\section{Comparison to the classical case}
\label{sec:classical}

In the classical case $\beta_{z}(t)=t \ln(t/\nu(z))$, in which $B$ becomes the information (or
Kullback-Leibler) divergence and $\Ecal$ is an exponential family with reference measure~$\nu$, the
function $\Bbar$, which, in the general case, is defined by means of an optimization problem, has an
explicit analytic expression:
\begin{equation*}
  \Bbar(u) = \ln\bigg(1 + \exp\Big(\sum_{z\in Z}\frac{u(z)}{\|u\|_{1}}\ln|u(z)|\Big)\bigg) = \ln\big(1 + \exp(\ol D(u))\big).
\end{equation*}
Thus, while an optimization problem has to be solved to evaluate the function $B(\cdot,\Ecal)$ at
some $P\in\Pcal(Z)$, the function $\Bbar$ can be evaluated more easily.

In the general case this is not true anymore.  However, the computational complexity of the
optimization problem~\ref{prob:Bbar} is still different from the complexity of the
problem~\ref{prob:B}.  To evaluate $\Bbar(u)$ at a single point
$u\in\Ncal\setminus\{0\}$, a problem of a similar kind as problem~\ref{prob:B}, but much smaller,
has to be solved: the solution is a pm in $\Pcal(\supp(u^{+}))$.  Moreover, as $\Fcal_{u}$ has
co-dimension one, $rB$-projections to $\Fcal_{u}$ can be computed by solving a one-dimensional
optimization problem (namely, $\Pi_{\Fcal_{u},P}$ minimizes $H(Q)$ for $Q\in P + \Rb u$).

In total, whether it is easier to attack problem~\ref{prob:Bbar} or~\ref{prob:B} may depend on the specific choice of the functions $\beta_{z}$ and~$f$.
For the classical case, \cite{Rauh11:Thesis} and~\cite{Rauh11:Finding_Maximizers} present many ideas how to attack problem~\ref{prob:bar-problem-KL}, many of which may generalize to problem~\ref{prob:Bbar}, depending on the choice of the functions~$\beta_{z}$.

Most importantly, the idea behind the definition of the function $\Bbar$ sheds light on the relation of the problems~\ref{prob:main-problem-KL} and~\ref{prob:bar-problem-KL}, which is rather opaque if one only looks at the definitions of the functions~$D$ and~$\ol D$.


\section*{Acknowledgement}
\small
This work was partially supported by
the Grant Agency of the
Czech Republic under Grant P202/10/0618 and the Research
Academy Leipzig.

\section*{Author contributions}

The first investigations were done by the second author in 2010, who also provided the correct notion of a Bregman family.  In 2012, both authors worked together to find a good definition for $\Bbar$ and to prove the equivalence of the global maximizers (Theorem~\ref{thm:equivalence}).  The project was delayed by the first author trying to find a proof of Conjecture~\ref{con:uniqueness-codim-one}.  
The first author added further results and completed the manuscript.


\bibliographystyle{IEEEtranSpers}
\bibliography{general}

\makecontacts

\end{document}